\documentclass[showpacs,preprintnumbers,twocolumn,superscriptaddress,aps,nofootinbib]{revtex4}
\usepackage{amssymb}
\usepackage[centertags]{amsmath}
\usepackage{txfonts}
\usepackage{comment}
\usepackage{bm}
\usepackage{color}
\usepackage{graphicx,graphics}
\usepackage{multirow}
\usepackage{float}
\usepackage{ulem}
\usepackage[colorlinks,citecolor=blue,linkcolor=blue]{hyperref}

\begin{document}

\title{Yield ratios of identified hadrons in p+p, p+Pb, Pb+Pb collisions at the Large Hadron Collider}

\author{Feng-lan Shao}
\affiliation{Department of Physics, Qufu Normal University, Shandong 273165, China}

\author{Guo-jing Wang}
\affiliation{Department of Physics, Qufu Normal University, Shandong 273165, China}

\author{Rui-qin Wang}
\affiliation{Department of Physics, Qufu Normal University, Shandong 273165, China}


\author{Hai-hong Li}
\affiliation{Department of Physics, Jining University, Shandong 273155, China}

\author{Jun Song}\email{songjun2011@jnxy.edu.cn}
\affiliation{Department of Physics, Jining University, Shandong 273155, China}

\begin{abstract}
Yield ratios of identified hadrons observed in high multiplicity p+p and p+Pb collisions at LHC show remarkable similarity with those in Pb+Pb collisions, indicating some important and universal underlying dynamics in hadron production for different quark gluon final states.  
	We use the quark combination model to explain the data of yield ratios in these three collision systems.
	The observed $p/\pi$ and $\Lambda/K_s^0$ can be reproduced simultaneously by quark combination, and these two ratios reflect the probability of baryon production at hadronization which is the same in light sector and strange sector and is roughly constant in p+p, p+Pb and Pb+Pb collision systems over three orders of magnitude in charged particle multiplicity.
	The data of $K_s^0/\pi$, $\Lambda/\pi$, $\Xi/\pi$ and $\Omega/\pi$ show a hierarchy behavior relating to the strangeness content, and are naturally explained by quark combination both in the saturate stage at high multiplicity and in the increase stage at moderate multiplicity.
	Our results suggest that the characteristic of quark combination is necessary in describing the production of hadrons in small systems created in p+p and p+Pb collisions. 

\end{abstract}

\maketitle

\section{Introduction}
Ultra-relativistic collisions of nucleon-nucleon, nucleon-nucleus and nucleus-nucleus create much different parton systems in size, e.g.,~ over three orders of magnitude if we use the charged particle multiplicity as a rough estimation of system size. 
Unexpectedly, recent ALICE experiments at LHC find a series of remarkable similarities for hadron production in high multiplicity p+p, p+Pb collisions and Pb+Pb collisions, e.g.,~long range angular correlations \cite{ridgeppJHEP2010,ridgepPbPLB2013}, flow-like patterns \cite{flow2014NPA}, enhanced strangeness \cite{xiomepPb,KLamXiOmgpp7TeV}, and enhanced baryon to meson ratios at soft transverse momenta \cite{piKpLampPb,lampPb16}.  These similarities invoke intensive discussions in literatures involving the mini-QGP or phase transition \cite{lfm11,Kw11,Bzdak13,Bozek13,prasad10,Avsar11}, multiple parton interactions \cite{MPI}, string overlap \cite{Bautista15,Bierlich2015} and color reconnection \cite{Velasquez13} at hadronization in small systems created in p+p and p+Pb collisions at LHC.

Yield ratios of identified hadrons are one kind of particularly important observables because of their sensitivity on the microscopic mechanism of hadron production at hadronization.  The data of $p/\pi$ and $\Lambda/K_s^0$ characterizing the relative baryon production and these of $K_s^0/\pi$, $\Lambda/\pi$, $\Xi/\pi$ and $\Omega/\pi$ characterizing the strangeness production in high multiplicity p+p and p+Pb collisions \cite{piKpLampPb,xiomepPb,KLamXiOmgpp7TeV} show remarkable similarity with those in Pb+Pb collisions \cite{piKpPbPb,Ks0LPbPb,XiOmgPbPb} at LHC, indicating some universal underlying dynamics in hadron production for different quark gluon final states.  
According to the report of Ref.\cite{KLamXiOmgpp7TeV}, popular event generators such as PYTHIA8 \cite{pythia8}, EPOS \cite{eposlhc}, DIPSY \cite{Bierlich2015,dipsy}, PHOJET \cite{phojet} and HERWIG \cite{herwig} can not consistently explain the data of above yield ratios, which suggests that some of new characteristic of hadronization beyond the traditional string/cluster fragmentation maybe play the role in p+p and p+Pb collisions. 

In this paper, we understand the experimental data of yields of pions, kaons, proton, $\Lambda$, $\Xi$ and $\Omega$ in Pb+Pb collisions, in high-multiplicity events of p+Pb collisions, and in that of p+p collisions at LHC in quark combination mechanism. 
Quark combination is an effective hadronization mechanism and has been adopted by various hadron production models and event generators such as AMPT \cite{ampt}, and has been used to explain the data of $e^++e^-$ annihilations and p+p reactions already in the 1970s \cite{Anisovich1973NPB,Bjorken1974PRD,Das77,xqb88} and the data of relativistic A+A collisions in recent years \cite{Zimanyi2000PLB,Greco2003PRL,Fries2003PRL,RCHwa2003PRC,FLShao2005PRC,co2006PRC,lzw2016AMPT,lzw2017AMPT}, especially the data of yields and momentum distributions of identified hadrons. 
Using two classes of yield ratios, i.e.,~ratios of baryons to mesons and ratios of strange and multi-strange hadrons to pions, we study the property of the baryon production probability in the above different collision systems at LHC and the hierarchy phenomenon in production of strange hadrons with different strange quanta, and we illustrate how these observed properties can be naturally explained by the quark combination.  

The rest of the paper is organized as follows.  In Sec.~II, we discuss the application of quark combination to different systems produced in p+p and p+Pb, and Pb+Pb collisions, and introduce the formulas of hadronic yields in quark combination model. In Sec.~III, we first give the global fit of experimental data of identified hadrons in p+p, p+Pb, and Pb+Pb collisions, and discuss the correlations of hadronic yields hidden in the experimental data by analyzing two classes of  yield ratios.  A summary is presented in Sec.~IV.

\section{quark combination and hadronic yield}

Quark combination mechanism (QCM) describes the formation of hadrons at hadronization by the combination of quarks and antiquarks neighboring in phase space. The mechanism assumes the effective absence of soft gluon quanta at hadronization and the effective degrees of freedom of QCD matter are quarks and antiquarks with constituent masses at hadronization. Application of quark combination to the bulk quark system produced in relativistic A+A collisions is natural in picture, and has good performance in explaining or reproducing the data of transverse momentum spectra, yields and longitudinal rapidity distributions for various identified hadrons \cite{Greco2003PRL,Fries2003PRL,RCHwa2003PRC, Zimanyi2000PLB,FLShao2005PRC,co2006PRC,sj2013,wrq14, RQWang2012PRC,CEShao2009PRC,RQWang2015PRC}. Application of quark combination to small systems created in $e^++e^-$ and p+p collisions is not popular because the string/cluster fragmentation is usually regarded as the standard recipe in these collisions. However, by treating the string or color-neutral cluster as a collection of quarks and antiquarks with constituent masses and then combining them to various hadrons, quark combination mechanism had also reproduced many experimental data of $e^++e^-$ and p+p collisions in early years \cite{xqb88,ZhaoJQ95,WQ96,SZG97}. 

In this paper, we concentrate on the yield properties of identified hadrons. One of the main features of quark combination in hadronic yield is that the flavor content of identified hadron determines its yield, not the mass of hadron as in statistical hadronization model. Therefore, we write the average yield of identified hadrons after the hadronization of a quark system with given numbers of quarks and antiquarks in the following form
\begin{eqnarray}
    \overline{N}_{B_i} &=& N_{B_i}^{(q)} P_{q_1 q_2 q_3 \rightarrow B_i} , \label{eq1}\\
    \overline{N}_{M_i} &=& N_{M_i}^{(q)} P_{q_1\bar{q}_2 \rightarrow M_i},
    \label{eq2}
\end{eqnarray}
where $N_{B_i}^{(q)} = N_{iter}\prod_f \prod_{i=1}^{n_{f,B_i}} (N_f-i+1)$ is the number of three specific flavor combinations being relevant for $B_i$ formation. $N_f$ is the number of quark of flavor $f$ in system just before hadronization.  $n_{f,B_i}$ is the number of valance quark $f$ contained in baryon $B_i$. $N_{iter}$ is the iteration factor taking to be 1, 3, and 6 for the cases of three identical flavor, two different flavors and three different flavors contained in a baryon, respectively.
We note that $N_{B_i}^{(q)}$ has correctly considered some necessary threshold effects for identified baryons. For example, $N_{\Omega^-}^{(q)}=N_s(N_s-1)(N_s-2)$ means that $\Omega^-$ can be only produced in events with strange quark number $N_s\geq 3$. 
$P_{q_1 q_2 q_3 \rightarrow B_i}$ is the average combination probability of $q_1 q_2 q_3\rightarrow B_i$. The meson formula is similarly defined. 
The number of specific-flavor quark antiquark pairs for $M_i$ formation is $N_{M_i}^{(q)} = \sum_{k} \omega_{k} \prod_f \prod_{i=1}^{n_{f,M_i,k}} (N_f-i+1)$ where $f$ runs over all flavors of quarks and antiquarks. This incorporates the case of mixed quark and antiquark flavors for some mesons, e.g.,~$\pi^0$ is composed by $u\bar{u}$ and $d\bar{d}$ with weights 1/2, respectively. Index $k$ runs over all channels of flavor mixing  and $\omega_{k}$ is the weight.    $n_{f,M_i,k}$ is the number of valance $f$ in $k$ channel, taking to be 1 or 0. $P_{q_1 \bar{q}_2} \rightarrow M_i$ is the average combination probability of $q_1 \bar{q}_2 \rightarrow M_i$.

The combination probabilities $P_{q_1 q_2 q_3 \rightarrow B_i}$ and $P_{q_1 \bar{q}_2 \rightarrow M_i }$ can be evaluated as 
\begin{eqnarray}
    P_{q_1 q_2 q_3 \rightarrow B_i} &=& C_{B_i} \frac{\overline{N}_B }{N_{qqq}}, \label{prob_B}\\
    P_{q_1 \bar{q}_2 \rightarrow M_i } &=& C_{M_i} \frac{ \overline{N}_M  }{ N_{q\bar{q}} }, \label{prob_M}
\end{eqnarray}
where $\overline{N}_{B} = \sum_{j} \overline{N}_{B_j}$ is the average number of total baryons and $\overline{N}_M =\sum_{j} \overline{N}_{M_j}$ is total mesons. $N_q = \sum_f N_f$ is the total quark number. $N_{qqq}=N_q(N_q-1)(N_q-2)$ is the total possible number of three quark combinations in baryon formation and $N_{q\bar{q}}=N_q N_{\bar{q}}$ is the total number of quark antiquark pairs in meson formation. 
 Considering the flavor independence of strong interactions, $\overline{N}_B/N_{qqq}$ is used to approximately denote the average probability of three quarks combining into a baryon and $C_{B_i}$ is the branch ratio to $B_i$ for a given flavor $q_1 q_2 q_3$ combination. Similarly, $\overline{N}_M/N_{q\bar{q}}$ is used to approximately denote the average probability of a quark and antiquark combining into a meson and $C_{M_i}$ is the branch ratio to $M_i$ for a given flavor $q_1\bar{q}_2$ combination.

Here we consider only the ground state $J^P=0^-,\,1^-$ mesons and $J^P=(1/2)^+,\,(3/2)^+$ baryons in flavor SU(3) group. For mesons
\begin{equation}
C_{M_j} =  \left\{
\begin{array}{ll}
{1}/{(1+R_{V/P})}~~~~~~~~   \textrm{for } J^P=0^-  \textrm{ mesons}  \\
{R_{V/P}}/{(1+R_{V/P})}~~~~         \textrm{for } J^P=1^-  \textrm{ mesons},
\end{array} \right.
\end{equation}
where the parameter $R_{V/P}$ represents the ratio of the $J^P=1^-$ vector mesons to the $J^P=0^-$ pseudoscalar mesons
of the same flavor composition; for baryons
\begin{equation}
C_{B_j} =  \left\{
\begin{array}{ll}
{R_{O/D}}/{(1+R_{O/D})}~~~~   \textrm{for } J^P=({1}/{2})^+  \textrm{ baryons}  \\
{1}/{(1+R_{O/D})}~~~~~~~~         \textrm{for } J^P=({3}/{2})^+  \textrm{ baryons},
\end{array} \right.
\end{equation}
except that $C_{\Lambda}=C_{\Sigma^0}={R_{O/D}}/{(1+2R_{O/D})},~C_{\Sigma^{*0}}={1}/{(1+2R_{O/D})},~C_{\Delta^{++}}=C_{\Delta^{-}}=C_{\Omega^{-}}=1$.
Here, $R_{O/D}$ stands for the ratio of the $J^P=(1/2)^+$ octet to the $J^P=(3/2)^+$ decuplet baryons of the same flavor composition. $R_{V/P}$ and $R_{O/D}$ are set to be constant 1.5 and 2, respectively. 

Some further explanations on the above yield formulas in QCM are needed. We directly write the hadronic yields as functions of quark numbers just before hadronization, which does not follow the usual procedure of hadron production in QCM \cite{Greco2003PRL,Fries2003PRL,RCHwa2003PRC,RQWang2012PRC} by starting from the discussion of the phase space (spatial and momentum) distributions of quarks and antiquarks and the detailed combination process locally happened. This is because that yield is a phase-space integrated quantity.  Whatever quarks and antiquarks populate in spatial and momentum space, finally they are all neighboringly combined into hadrons, and therefore the quark numbers are major quantities surviving the phase space integration and serve as the main control parameter on hadronic yields. Here, we neglect in yield formulas the possible influence of quark phase space distribution, in particular the difference between light and strange quarks. For the detailed discussions on approximations and/or assumptions underlying in hadronic yield in QCM, one can refer to Ref.~\cite{RQWang2012PRC}.

We should also consider the fluctuation of quark numbers for the quark system produced in collisions at fixed collisional energy, 
\begin{equation}
	\langle N_{h_i} \rangle =  \sum_{\{N_{q_j}, N_{\bar{q}_j}\}} \mathcal{P}({\{N_{q_j}, N_{\bar{q}_j}\}}) \, \overline{N}_{h_i},
    \label{nh_pdf}
\end{equation}
where $\mathcal{P}({\{N_{q_j}, N_{\bar{q}_j}\}})$ is the distribution of quark numbers and antiquark numbers. In this paper, we adopt Poisson distribution as a rough approximation,  i.e.,~$\mathcal{P}({\{N_{q_j}, N_{\bar{q}_j}\}})= \prod_{f}^{u,d,s} \text{Pois}(N_f; \langle N_f\rangle)$, and take zero baryon chemical potential by $N_{f}=N_{\bar{f}}$.
Effects of quark number distribution and the limitation of Poisson approximation are discussed in the following section.

Including the decay contributions, we get the yields of identified hadrons in the final state,
\begin{equation}
\langle N_{h_i}^f\rangle = \langle N_{h_i}\rangle+ \sum_{j\not=i} Br(h_j\to h_i) \langle N_{h_j}\rangle,
\end{equation}
where we use the superscript $f$ to denote the results for the final hadrons. $Br(h_j\to h_i)$ is the branch ratio of hadron $h_j$ decaying into $h_i$, and is taken from the Particle Data Group \cite{PDG2014}.
In the following text, we take the corresponding decay contribution/correction according to the experiments. 

\section{Results and discussions}

In this section, we apply the formulas of the hadronic yields in QCM obtained above to explain the yield densities of identified hadrons measured in Pb+Pb collisions at $\sqrt{s_{NN}}=2.76$ TeV and p+Pb collisions at $\sqrt{s_{NN}}=5.02$ TeV. 
We emphasize that applying the quark combination to a finite rapidity window is reasonable. This is because quark combination has the local property and the rapidity shift between the quarks and the hadron they form is usually small. Therefore, we can set quark numbers to be those in midrapidity region and obtain hadronic yields that can be compared to experimental data.
We first present the global fit of the yield densities of pions, kaons, protons, $\Lambda$, $\Xi^-$ and $\Omega^-$ at midrapidity, and then put particular emphasis on the discussion of baryon production probability via $p/\pi$ and $\Lambda/K_s^0$ ratios and on the production of strange and multi-strange hadrons relative to pions. 

\subsection{Yield densities of identified hadrons}
\label{subsec1}

The main inputs of our model are the average rapidity densities of quark numbers $\langle \mathrm{d}N_u/\mathrm{d}y \rangle$, $\langle \mathrm{d}N_d/\mathrm{d}y \rangle$, and $\langle \mathrm{d}N_s/\mathrm{d}y \rangle$ before hadronization. We use the strangeness suppression factor $\lambda_s = \langle \mathrm{d}N_s/\mathrm{d}y \rangle / \langle \mathrm{d} N_u/\mathrm{d}y \rangle$ to denote the production suppression of strange quarks relative to light quarks, and use $\langle \mathrm{d}N_q/\mathrm{d}y \rangle = \langle \mathrm{d}N_u/\mathrm{d}y \rangle + \langle \mathrm{d}N_d/\mathrm{d}y \rangle + \langle \mathrm{d}N_s/\mathrm{d}y \rangle $ to characterize the size of quark system at midrapidity.
Considering the isospin symmetry $\langle \mathrm{d}N_u/\mathrm{d}y \rangle = \langle \mathrm{d}N_d/\mathrm{d}y \rangle$, inputs on quark numbers are reduced to $\langle \mathrm{d}N_q/\mathrm{d}y \rangle$ and $\lambda_s$.

One important ingredient in our model which is not determined yet in previous section is the average baryon number $\overline{N}_B$ and average meson number $\overline{N}_M$ at given quark numbers and antiquark numbers. Unitarity of hadronization requires that all quarks and antiquarks are combined into hadrons, i.e.,~$\overline{N}_M + 3 \overline{N}_B = N_q$.  In our previous work \cite{sj2013}, we have studied the properties of $\overline{N}_B$ and $\overline{N}_M$ in the case of large quark numbers, and obtained $R_{B/M} \equiv \overline{N}_B/\overline{N}_M\approx 1/12$ at zero baryon quantum number density by reproducing the observed yield ratios in relativistic heavy ion collisions. Since in this paper we also fit the data of hadronic yields in the small system produced in p+Pb collisions, the baryon/meson competition factor $R_{B/M}$ is regarded as a parameter. We will study this point by using the data of $p/\pi$ and $\Lambda/K_s^0$ yield ratios in p+p, p+Pb and Pb+Pb collisions at LHC. 

Table \ref{tab1} shows the values of $\langle N_q \rangle$, $\lambda_s$ and $R_{B/M}$ obtained by fitting the yield data in Pb+Pb collisions at $\sqrt{s_{NN}}=2.76$ TeV, and these in p+Pb collisions at $\sqrt{s_{NN}}=5.02$ TeV. The strangeness suppression factor $\lambda_s$ is saturated in central and semi-central Pb+Pb collisions and decreases in peripheral Pb+Pb collisions and p+Pb collisions. The change of $\lambda_s$ is consistent with calculations of Lattice QCD \cite{lsLQCD06,lqcd06cbs} in grand-canonical ensemble and the parameter value in PYTHIA \cite{pythia8}.
 We note that the change of strangeness with respect to system size was also addressed by the statistical/thermal model \cite{shmLHC09,shm17} using the (grand-)canonical ensemble, which can qualitatively explain the data of two classes of yield ratios discussed in this paper \cite{piKpLampPb}. 

The baryon/meson competition factor $R_{B/M}$ increases slightly with the decrease of total quark number (or system size).
We note that final-state hadronic rescatterings in heavy ion collisions can slightly increase the pion yield and/or decrease proton yield by baryon-antibaryon annihilation reactions \cite{bass2000,shc15}.
Therefore, the extracted $R_{B/M}$ is, more or less, influenced by the final-state hadronic rescatterings. 
Since the relative magnitude of $R_{B/M}$ change is less than 9\% over three orders of magnitude in multiplicity, we neglect the discussion of final state effects in the following text which can not change our conclusion. 

\begin{table}[htb]
    \caption{The input values of $\lambda_s$, $\langle \mathrm{d}N_q/\mathrm{d}y \rangle$ at midrapidity and $R_{B/M}$ in Pb+Pb collisions at $\sqrt{s_{NN}}=2.76$ TeV, and these in p+Pb collisions at $\sqrt{s_{NN}}=5.02$ TeV.}
	\begin{tabular}{ p{2.2cm}|*{3}{ p{1.5cm} } }
	\toprule
                    &$\langle \mathrm{d}N_q/\mathrm{d}y \rangle$		&$\lambda_s$		&$R_{B/M}$ \\ \colrule
	PbPb 0-5\%		&1720						&0.405				&1/11.9 \\	
	PbPb 5-10\%		&1440						&0.410				&1/11.9 \\	
	PbPb 10-20\%	&1100						&0.410				&1/11.5 \\	
	PbPb 20-30\%	&750						&0.410				&1/11.5 \\	
	PbPb 30-40\%	&482						&0.412				&1/11.5 \\	
	PbPb 40-50\%	&288						&0.412				&1/11.5 \\	
	PbPb 50-60\%	&170						&0.410				&1/11.4 \\	
	PbPb 60-70\%	&84							&0.380				&1/11.4 \\	
	PbPb 70-80\%	&41							&0.365				&1/11.2 \\	
	PbPb 80-90\%	&15							&0.35				&1/11.2 \\	
	\colrule
	pPb 0-5\%		&52							&0.36				&1/11.0 \\	
	pPb 5-10\%		&41.5						&0.36				&1/11.0 \\	
	pPb 10-20\%		&35							&0.356				&1/11.0 \\	
	pPb 20-40\%		&26.5						&0.355				&1/11.0 \\	
	pPb 40-60\%		&18							&0.342				&1/11.0 \\	
	pPb 60-80\%		&11.2						&0.33				&1/11.0 \\	
	pPb 80-100\%	&4.7						&0.33				&1/11.0 \\	
	\botrule
\end{tabular}
	\label{tab1}
\end{table}

\begin{table*}[!htb]
\renewcommand{\arraystretch}{1.}
    \caption{Rapidity densities $\langle \mathrm{d}N/\mathrm{d}y\rangle$ of different hadrons at midrapidity in different centralities in Pb+Pb collisions at $\sqrt{s_{NN}}=2.76$ TeV. The data are from Refs. \cite{piKpPbPb,Ks0LPbPb,XiOmgPbPb}.}
\begin{tabular}{@{}c|cc|cc|cc|cc|cc@{}}
\toprule
 &\multicolumn{2}{c|}{0-5\% (0-10\%)}  &\multicolumn{2}{c|}{5-10\% (0-10\%)}  &\multicolumn{2}{c|}{10-20\%}  &\multicolumn{2}{c|}{20-30\% (20-40\%)}  &\multicolumn{2}{c}{30-40\% (20-40\%)}   \\
\raisebox{2.0ex}[0pt]{Centrality} &data  &model    &data  &model    &data  &model    &data  &model     &data  &model      \\
\colrule
$\pi^+$        &$733\pm54$     &698      &$606\pm42$     &583      &$455\pm31$   &443    &$307\pm20$     &296    &$201\pm13$     &194    \\
$\pi^-$        &$732\pm52$     &698      &$604\pm42$     &583      &$453\pm31$   &443    &$306\pm20$     &296    &$200\pm13$     &194     \\
$K^+$          &$109\pm9$      &108      &$91\pm7$       &91.2     &$68\pm5$     &69.2   &$46\pm4$       &46.2   &$30\pm2$       &30.4      \\
$K^-$          &$109\pm9$      &108      &$90\pm8$       &91.2     &$68\pm6$     &69.2   &$46\pm4$       &46.2   &$30\pm2$       &30.4      \\
$K_S^0$        &$110\pm10$     &104      &$90\pm6$       &88	   &$68\pm5$     &66.8   &$(39\pm3)$     &(37.6)   &$(39\pm3)$     &(35.8)      \\
$p$            &$34\pm3$       &33.2     &$28\pm2$       &27.6     &$21.0\pm1.7$ &21.6   &$14.4\pm1.2$   &14.4   &$9.6\pm0.8$    &9.48     \\
$\bar p$       &$33\pm3$       &33.2     &$28\pm2$       &27.6     &$21.1\pm1.8$ &21.6   &$14.5\pm1.2$   &14.4   &$9.7\pm0.8$    &9.48     \\
$\Lambda$      &$26\pm3$       &26.4     &$22\pm2$       &22.2     &$17\pm2$     &17.2   &$(10\pm1)$     &(9.8)   &$(10\pm1)$     &(9.34)     \\
$\Xi^-$        &$(3.34\pm0.30)$&(3.83) &$(3.34\pm0.30)$&(3.80)     &$2.53\pm0.22$&2.76 &$(1.49\pm0.13)$&(1.55)&$(1.49\pm0.13)$&(1.49)  \\
$\bar\Xi^+$    &$(3.28\pm0.29)$&(3.83) &$(3.28\pm0.29)$&(3.80)     &$2.51\pm0.23$&2.76 &$(1.53\pm0.13)$&(1.55)&$(1.53\pm0.13)$&(1.49)  \\
$\Omega^-$     &$(0.58\pm0.13)$&(0.51) &$(0.58\pm0.13)$&(0.51)     &$0.37\pm0.09$&0.37 &$(0.23\pm0.04)$&(0.21)&$(0.23\pm0.04)$&(0.20)  \\
$\bar\Omega^+$ &$(0.60\pm0.14)$&(0.51) &$(0.60\pm0.14)$&(0.51)     &$0.40\pm0.09$&0.37 &$(0.25\pm0.04)$&(0.21)&$(0.25\pm0.04)$&(0.20)  \\
\botrule
\end{tabular}
\begin{tabular}{@{}c|cc|cc|cc|cc|cc@{}}
\toprule
  &\multicolumn{2}{c|}{40-50\% (40-60\%)}  &\multicolumn{2}{c|}{50-60\% (40-60\%)}  &\multicolumn{2}{c|}{60-70\% (60-80\%)}  &\multicolumn{2}{c|}{70-80\% (60-80\%)}  &\multicolumn{2}{c}{80-90\%} \\
\raisebox{2.0ex}[0pt]{Centrality} &data  &model    &data  &model    &data  &model    &data  &model     &data  &model         \\
\colrule
$\pi^+$        &$124\pm8$        &116     &$71\pm5$         &68.5    &$37\pm2$         &34.4    &$17.1\pm1.1$     &16.9    &$6.6\pm0.4$    &6.45    \\
$\pi^-$        &$123\pm8$        &116     &$71\pm4$         &68.5    &$37\pm2$         &34.4    &$17.0\pm1.1$     &16.9    &$6.6\pm0.4$    &6.45    \\
$K^+$          &$18.3\pm1.4$     &18.1    &$10.2\pm0.8$     &10.6    &$5.1\pm0.4$      &4.9     &$2.3\pm0.2$      &2.32     &$0.85\pm0.08$  &0.837  \\
$K^-$          &$18.1\pm1.5$     &18.1    &$10.2\pm0.8$     &10.6    &$5.1\pm0.4$      &4.9     &$2.3\pm0.2$      &2.32     &$0.86\pm0.09$  &0.837  \\
$K_S^0$        &$(14\pm1)$       &(14.2)  &$(14\pm1)$       &(13.3)  &$(3.9\pm0.2)$    &(3.6)   &$(3.9\pm0.2)$    &(3.19)   &$0.85\pm0.09$  &0.806  \\
$p$            &$6.1\pm0.5$      &5.7     &$3.6\pm0.3$      &3.4     &$1.9\pm0.2$      &1.74     &$0.90\pm0.08$    &0.873    &$0.36\pm0.04$  &0.326        \\
$\bar p$       &$6.2\pm0.5$      &5.7     &$3.7\pm0.3$      &3.4     &$2.0\pm0.2$      &1.74     &$0.93\pm0.09$    &0.873    &$0.36\pm0.04$  &0.326               \\
$\Lambda$      &$(3.8\pm0.4)$    &(3.7)   &$(3.8\pm0.4)$    &(3.5)   &$(1.0\pm0.1)$    &0.96     &$(1.0\pm0.1)$    &(0.888)   &$0.21\pm0.03$  &0.222  \\
$\Xi^-$        &$(0.53\pm0.05)$  &(0.58)  &$(0.53\pm0.05)$  &(0.55)  &$(0.124\pm0.012)$&(0.141) &$(0.124\pm0.012)$&(0.125) &$---$          &0.029  \\
$\bar\Xi^+$    &$(0.54\pm0.05)$  &(0.58)  &$(0.54\pm0.05)$  &(0.55)  &$(0.120\pm0.011)$&(0.141) &$(0.120\pm0.011)$&(0.125) &$---$          &0.029  \\
$\Omega^-$     &$(0.087\pm0.019)$&(0.080) &$(0.087\pm0.019)$&(0.075) &$(0.015\pm0.005)$&(0.0176) &$(0.015\pm0.005)$&(0.015) &$---$          &0.0034  \\
$\bar\Omega^+$ &$(0.082\pm0.018)$&(0.080) &$(0.082\pm0.018)$&(0.075) &$(0.017\pm0.005)$&(0.0176) &$(0.017\pm0.005)$&(0.015) &$---$          &0.0034  \\
\botrule
\end{tabular}
\label{tab2}
\end{table*}

\begin{table*}[!htb]
\renewcommand{\arraystretch}{1.}
    \caption{Rapidity densities $\langle \mathrm{d}N/\mathrm{d}y \rangle$ of different hadrons in midrapidity in different centralities in p+Pb collisions at $\sqrt{s_{NN}}=5.02$ TeV. 
The data are from Refs. \cite{piKpLampPb,xiomepPb}.}
\begin{tabular}{@{}c|cc|cc|cc@{}}
\toprule
 &\multicolumn{2}{c|}{0-5\%}  &\multicolumn{2}{c|}{5-10\%}  &\multicolumn{2}{c}{10-20\%}    \\
\raisebox{2.0ex}[0pt]{Centrality} &data  &model    &data  &model    &data  &model                \\
\colrule
$\pi^++\pi^-$           &$40.7568\pm6.1841$              &42.8    &$33.1034\pm4.8391$              &34.18   &$28.0460\pm3.8404$              &28.9      \\
$K^++K^-$               &$5.8450\pm0.9632$               &5.80    &$4.6414\pm0.7203$               &4.621   &$3.8916\pm0.5815$               &3.874       \\
$K_S^0$                 &---                             &2.802   &---                             &2.229   &---                             &1.867       \\
$p+\bar p$              &$2.2780\pm0.3637$               &2.259   &$1.8521\pm0.2862$               &1.805   &$1.5742\pm0.2364$               &1.531     \\
$\Lambda$               &$0.8143\pm0.1745$               &0.795   &$0.6521\pm0.1360$               &0.634   &$0.5465\pm0.1097$               &0.533          \\
$\Xi^-+\bar\Xi^+$       &$0.2354\pm0.0020\pm0.0161$      &0.221   &$0.1861\pm0.0016\pm0.0138$      &0.176   &$0.1500\pm0.0010\pm0.0112$      &0.147      \\
$\Omega^-+\bar\Omega^+$ &$0.0260\pm0.0011\pm0.0034$      &0.0261  &$0.0215\pm0.0008\pm0.0029$      &0.0209  &$0.0167\pm0.0006\pm0.0022$      &0.0174     \\
\botrule
\end{tabular}
\begin{tabular}{@{}c|cc|cc@{}}
\toprule
  &\multicolumn{2}{c|}{20-40\%}  &\multicolumn{2}{c}{40-60\%}     \\
\raisebox{2.0ex}[0pt]{Centrality} &data  &model    &data  &model            \\
\colrule
$\pi^++\pi^-$           &$21.7647\pm2.9790$              &21.97    &$15.2671\pm2.1827$                  &15.08    \\
$K^++K^-$               &$2.9600\pm0.4325$               &2.909   &$2.0264\pm0.3026$                   &1.918       \\
$K_S^0$                 &---                             &1.402   &---                                 &0.924       \\
$p+\bar p$              &$1.2211\pm0.1788$               &1.158   &$0.8594\pm0.1299$                   &0.801     \\
$\Lambda$               &$0.4169\pm0.0827$               &0.402   &$0.2835\pm0.0573$                   &0.268       \\
$\Xi^-+\bar\Xi^+$       &$0.1100\pm0.0006\pm0.0085$      &0.111   &$0.0726\pm0.0006\pm0.0065$          &0.0707    \\
$\Omega^-+\bar\Omega^+$ &$0.0120\pm0.0005\pm0.0016$      &0.0130  &$0.0072\pm0.0003\pm0.0010$          &0.00793      \\
\botrule
\end{tabular}
\begin{tabular}{@{}c|cc|cc@{}}
\toprule
   &\multicolumn{2}{c|}{60-80\%}    &\multicolumn{2}{c}{80-100\%}  \\
\raisebox{2.0ex}[0pt]{Centrality} &data  &model    &data  &model                   \\
\colrule
$\pi^++\pi^-$           &$9.4558\pm1.3792$                   &9.8314    &$4.1539\pm0.7655$                   &4.1305       \\
$K^++K^-$               &$1.2161\pm0.1821$                   &1.1404    &$0.5121\pm0.0960$                   &0.4660         \\
$K_S^0$                 &---                                 &0.5599    &---                                 &0.2289         \\
$p+\bar p$              &$0.5326\pm0.0822$                   &0.4906    &$0.2225\pm0.0430$                   &0.2078       \\
$\Lambda$               &$0.1677\pm0.0350$                   &0.1614    &$0.0643\pm0.0155$                   &0.0663           \\
$\Xi^-+\bar\Xi^+$       &$0.0398\pm0.0004\pm0.0031$          &0.0425    &$0.0143\pm0.0003\pm0.0015$          &0.0170        \\
$\Omega^-+\bar\Omega^+$ &$0.0042\pm0.0002\pm0.0006$          &0.0048    &$0.0013\pm0.0003\pm0.0003$          &0.0019       \\
\botrule
\end{tabular}
\label{tab3}
\end{table*}

Table \ref{tab2} and \ref{tab3} show our results of yield densities of identified hadrons in Pb+Pb collisions at $\sqrt{s_{NN}}=2.76$ TeV, and these in p+Pb collisions at $\sqrt{s_{NN}}=5.02$ TeV.
The experimental data of Pb+Pb collisions are from Refs. \cite{piKpPbPb,Ks0LPbPb,XiOmgPbPb} and these of p+Pb collisions are from Refs.\cite{piKpLampPb,xiomepPb}. The statistical uncertainties and systematic uncertainties of experimental data are also shown in the two tables, if available. 
In Table \ref{tab2}, due to the centrality interval difference in experiments for different species of hadrons, we use parentheses to differentiate in the same column.
Our results are rescaled according to the number of nucleon participants in order to compare with experimental data in slightly different centralities in the same column. 

We can see from Table \ref{tab2} that the results of QCM globally agree well with the available experimental data from central to peripheral Pb+Pb collisions. In p+Pb collisions, as shown in Table \ref{tab3}, we also see the well agreement between our results and experimental data in both high multiplicity events and low multiplicity events, considering the available statistical and systematic uncertainties.

\subsection{Baryon to meson ratios}

\begin{figure}[!htbp]
  \centering
  \includegraphics[width=\linewidth]{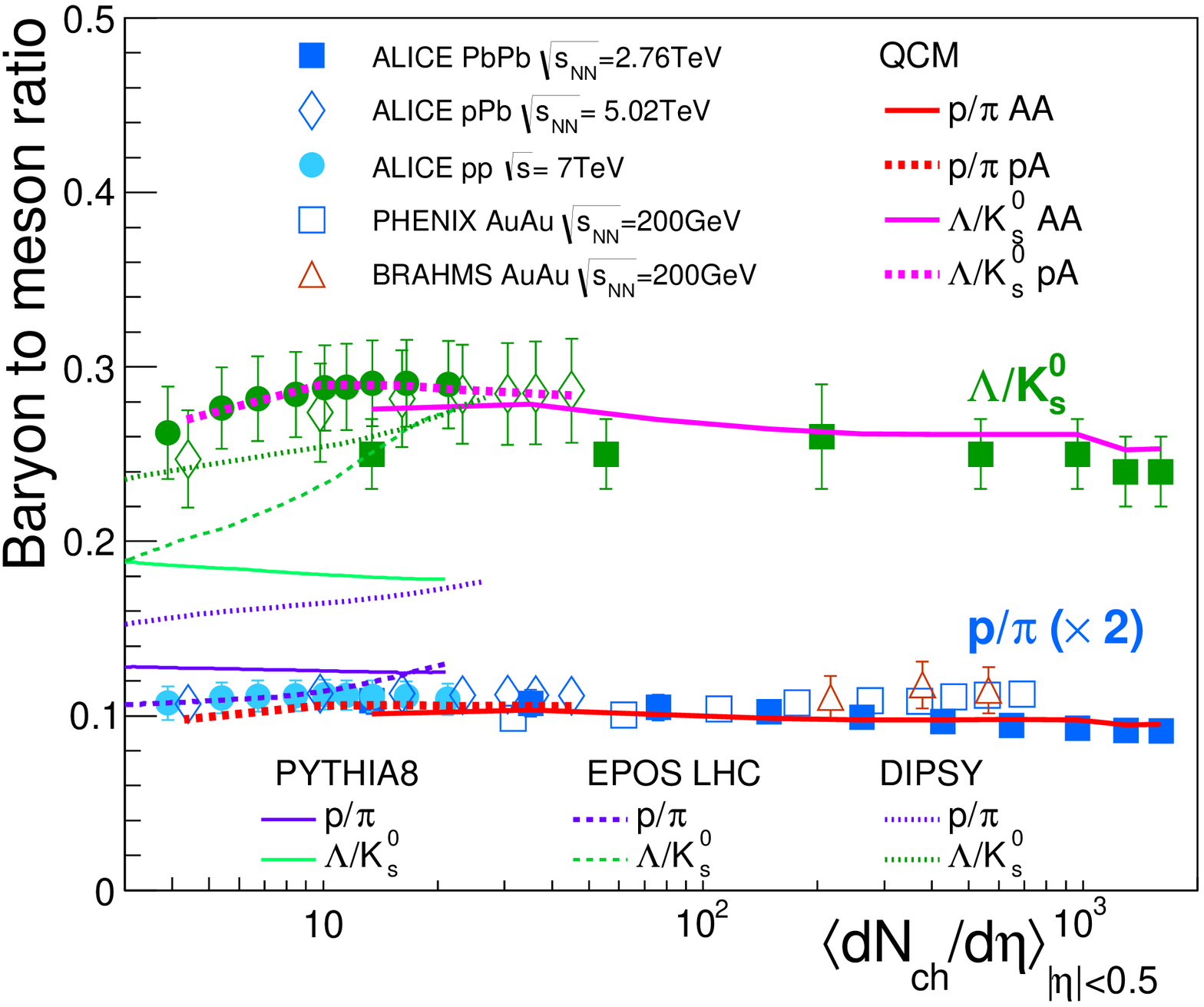}\\
	\caption{(Color online) Yield ratios $p/\pi=(p+\bar{p})/(\pi^{+}+\pi^{-})$ and $\Lambda/K_s^0=(\Lambda+\bar{\Lambda})/2K_s^0$ as the function of charged-particle multiplicity density at midrapidity. The experimental data of p+p, p+Pb, and Pb+Pb collisions are from Refs.~\cite{KLamXiOmgpp7TeV,piKpLampPb,piKpPbPb,Ks0LPbPb}, and data of Au+Au collisions at $\sqrt{s_{NN}}=200$ GeV are from Refs.~\cite{star034909,brahms202301,phenix034909}. Results of PYTHIA8, EPOS and DIPSY are taken from Ref.~\cite{KLamXiOmgpp7TeV}.}
  \label{fig1}
\end{figure}

Yield ratios are more sensitive observables to probe the hadron production mechanism. 
Here we first study two yield ratios, $p/\pi=(p+\bar{p})/(\pi^{+}+\pi^{-})$ and $\Lambda/K_s^0=(\Lambda+\bar{\Lambda})/2K_s^0$, which can effectively reflect some important dynamics of baryon production at hadronization. 
Fig.~\ref{fig1} shows the $p/\pi$ and $\Lambda/K_s^0$ at midrapidity as the function of the charged-particle multiplicity. The experimental data of Pb+Pb collisions at $\sqrt{s_{NN}}=2.76$ TeV \cite{piKpPbPb,Ks0LPbPb}, p+Pb collisions at $\sqrt{s_{NN}}=5.02$ TeV \cite{piKpLampPb}, p+p collisions at $\sqrt{s}=7$ TeV \cite{KLamXiOmgpp7TeV}, as well as those in Au+Au collisions at RHIC energies \cite{star034909,brahms202301,phenix034909} are shown in the figure. Some features exhibited from these experimental data are quite striking. For $p/\pi$ ratio, we see that the data of p+p collisions (full circles) are consistent with those of p+Pb collisions (open diamond) at the same multiplicities, and, in particular, they are almost independent of the multiplicity density as $\langle \mathrm{d}N_{ch}/\mathrm{d}\eta \rangle >3$. We also see with some surprise that the data of Pb+Pb collisions are weakly dependent on the charged-particle multiplicity and they are consistent with those in p+p and p+Pb collisions in magnitude. For $\Lambda/K_s^0$ ratio, the data of p+p collisions are also consistent with those of p+Pb collisions, and they increase slowly with the charged-particle multiplicity in small multiplicity events and saturate as $\langle \mathrm{d}N_{ch}/\mathrm{d}\eta \rangle\gtrsim 8$. The data of $\Lambda/K_s^0$ ratio in Pb+Pb collisions, shown as full squares, are also almost independent of the charged-particle multiplicity. The center values of peripheral collisions ($14\lesssim \langle \mathrm{d}N_{ch}/\mathrm{d}\eta\rangle \lesssim 50$) are slightly lower than those of p+p and p+Pb collisions at the similar charged-particle multiplicities, but within available statistical and systemic uncertainties they can be still regarded to be consistent with each other. In addition, the small change of $\Lambda/K_s^0$ ratios in the whole $\langle \mathrm{d}N_{ch}/\mathrm{d}\eta\rangle$ range can be partially attributed to the change of strangeness as discussed below in our method. Therefore, from the data of $p/\pi$ and $\Lambda/K_s^0$ in p+p, p+Pb and Pb+Pb collisions, we can say that we observe a stable baryon production probability in different system sizes, which is an indication of some universal properties in non-perturbative stage (for final-state parton system) evolved in different high energy collisions. 

The thick solid lines and dashed lines in Fig.~\ref{fig1} are results of QCM in Pb+Pb collisions and these of p+Pb collisions, respectively. The slight difference between our results in Pb+Pb collisions and these in p+Pb collisions in the overlap range $ 15 \lesssim \langle \mathrm{d}N_{ch}/\mathrm{d}\eta\rangle \lesssim 50 $ is due to the slightly different values of parameters $\lambda_s$ and $R_{B/M}$ which are listed in Table \ref{tab1} and are separately obtained by the best global fit of the data of hadronic yields in two collision systems. 
Results of p+p collisions are not given in Fig.~\ref{fig1} because of the following reasons. The experimental data of the absolute yields of identified hadrons in different centralities (or event activities) in p+p collisions at $\sqrt{s}=7$ TeV are unavailable at present. We can not make a global fit as did in p+Pb collisions in Sec.~\ref{subsec1} to directly give the results of yield ratios in p+p collisions. In addition, yield ratios in QCM, as explained in the following text, are mainly influenced by two parameters $\lambda_s$ and $R_{B/M}$ which are kinds of intensive quantities. They are not expected to change much in p+p and p+Pb collisions at similar charged particle multiplicity, which is indicated by, on the one hand, our fitting results of peripheral Pb+Pb collisions and central p+Pb collisions in Table \ref{tab1} and by, on the other hand, the similarity of the data in p+p and p+Pb collisions. Therefore, results in p+p collisions in QCM are expected to be almost the same as those in p+Pb collisions at similar charged particle multiplicity.

Our results of two yield ratios can be expressed as the analytic forms following a certain approximation, which are helpful to understand how QCM works.  
As quark numbers are not small, we can expand the hadronic yield formulas in Eqs.~(\ref{eq1}) and (\ref{eq2}) as the Taylor series at the event-average of quark numbers
\begin{equation}
\begin{split}
    \overline{N}_{h_i} = &\left. \overline{N}_{h_i} \right|_{\langle N_{f} \rangle } + \sum_{f_1} \left. \frac{\partial \overline{N}_{h_i} }{ \partial N_{f_1}} \right|_{\langle N_{f} \rangle }\delta N_{f_1} \\
    + & \frac{1}{2} \sum_{f_1, f_2} \left. \frac{\partial^2 \overline{N}_{h_i} }{ \partial N_{f_1} \partial N_{f_2}} \right|_{\langle N_{f} \rangle }\delta N_{f_1} \delta N_{f_2} + \mathcal{O}\left(\delta N_f^3 \right),
\end{split}
\end{equation}
where indexes $f_1,f_2$ run over all quark and antiquark flavors and $\delta N_{f_1} = N_{f_1} - \langle N_{f_1} \rangle$. The subscript $\langle N_{f} \rangle$ denotes the evaluation at the averaged quark numbers.
Substituting the above expansion into Eq.~(\ref{nh_pdf}), we get
\begin{equation}
    \langle N_{h_i} \rangle = \overline{N}_{h_i} + \frac{1}{2} \sum_{f_1,f_2} \frac{\partial^2 \overline{N}_{h_i} }{ \partial N_{f_1} \partial N_{f_2}} \left\langle \delta N_{f_1} \delta N_{f_2} \right\rangle + \mathcal{O}\left(\langle \delta N_f^3 \rangle\right),
\end{equation}
where we have dropped the subscript for convenience. We can see that quark number distribution influences the hadronic yield by the quark number fluctuations and correlations at second and higher orders. This influence is weak as quark numbers are not small. 
We take the leading term as the approximation of hadronic yield and obtain very simple yield formulas, e.g.,~ for hadrons without mixing flavors, 
\begin{eqnarray}
    \langle N_{B_i} \rangle &\approx C_{B_i} \frac{\lambda_s^{n_{s,B_i}}}{\left(2+\lambda_s\right)^3} \langle N_B \rangle, 
    \label{nbi_approx} \\
    \langle N_{M_i} \rangle &\approx C_{M_i} \frac{\lambda_s^{n_{s,M_i}}}{\left(2+\lambda_s\right)^2} \langle N_M \rangle, 
    \label{nmi_approx}
\end{eqnarray}
where $n_{s,B_i}$ ($n_{s,M_i}$) is the number of strange quarks and/or strange antiquarks in baryon (meson) $i$ in quark model. Using these formula and taking into account decay contribution, we get 
\begin{equation}
\begin{split}
	\frac{p + \bar{p}}{\pi^+ + \pi^-} \approx 4R_{B/M} \, \Big[ &(2+\lambda_s)(2.43+0.8\lambda_s+0.292\lambda_s^2) \\
										 &+(\frac{8}{3}+2.02\lambda_s+\frac{4}{3}\lambda_s^2)R_{B/M} \Big]^{-1}, 
\end{split}\label{Rppion}
\end{equation}

\begin{equation}
	\frac{\Lambda + \bar{\Lambda}}{2K_S^0} \approx \frac{7.74}{(2+\lambda_s)(1+0.2\lambda_s)} R_{B/M},   \label{lamkso}                                
\end{equation}
which show explicit dependence on  parameters $R_{B/M}$ and $\lambda_s$ and the correlation between two ratios.  Here, only strong and electromagnetic decays are included following the experimental corrections. With parameter values $\lambda_s \sim 0.34 - 0.42$ and $R_{B/M} \sim 1/11 - 1/12$, $\Lambda/K_s^0$ ratio is 0.25-0.28 and $p/\pi$ ratio 0.048-0.054, which are very close to our numerical results (thick lines in Fig.~\ref{fig1}) and are consistent with the experimental observations. 
Actually, one can use Eqs.~(\ref{nbi_approx}) and (\ref{nmi_approx}) to directly fit the yield data of various identified hadrons shown in Tables \ref{tab2} and \ref{tab3} and almost equally well explain the experimental data. Therefore, Eqs.~(\ref{Rppion}) and (\ref{lamkso}) correctly reveal the most basic physics among two ratios of baryons to mesons.

Here we argue that the self-consistent explanation of the data of $p/\pi$ and $\Lambda/K_s^0$ is crucial test for the hadronization of small parton systems $\mathrm{d}N_{ch}/\mathrm{d}\eta \lesssim 50$.  We also show results of PYTHIA8 \cite{pythia8}, DIPSY \cite{dipsy} and EPOS \cite{eposlhc} in Fig.~\ref{fig1}, which are taken from Ref.~\cite{KLamXiOmgpp7TeV}. The first two models adopt string fragmentation mechanism and the last adopts string-model-like parameterization for hadronic structure of cut Pomerons at hadronization.  We see that PYTHIA8 (thin solid lines) gives weakly changed baryon to meson ratios in small system but obviously underestimates the $\Lambda/K_s^0$ ratios with slight over-estimation on the $p/\pi$ ratio. DIPSY (thin dotted lines) well explains the $\Lambda/K_s^0$ ratios but overestimates the $p/\pi$ ratio. EPOS (thin dashed lines) well explains the $p/\pi$ ratio but underestimates the $\Lambda/K_s^0$ at $\langle \mathrm{d}N_{ch}/\mathrm{d}\eta \rangle\lesssim 10$, and it predicts too rapid increase of $\Lambda/K_s^0$ with the charged-particle multiplicity. The purpose of presenting these model results is to suggest that, as discussed above, if we use the quark combination instead of traditional string fragmentation to describe the hadron production (in particular baryon production) at hadronization, we can naturally understand the observed baryon to meson ratios. 

\subsection{Yield ratios of strange and multi-strange hadrons to pions}

\begin{figure}[!htbp]
  \centering
  \includegraphics[width=\linewidth]{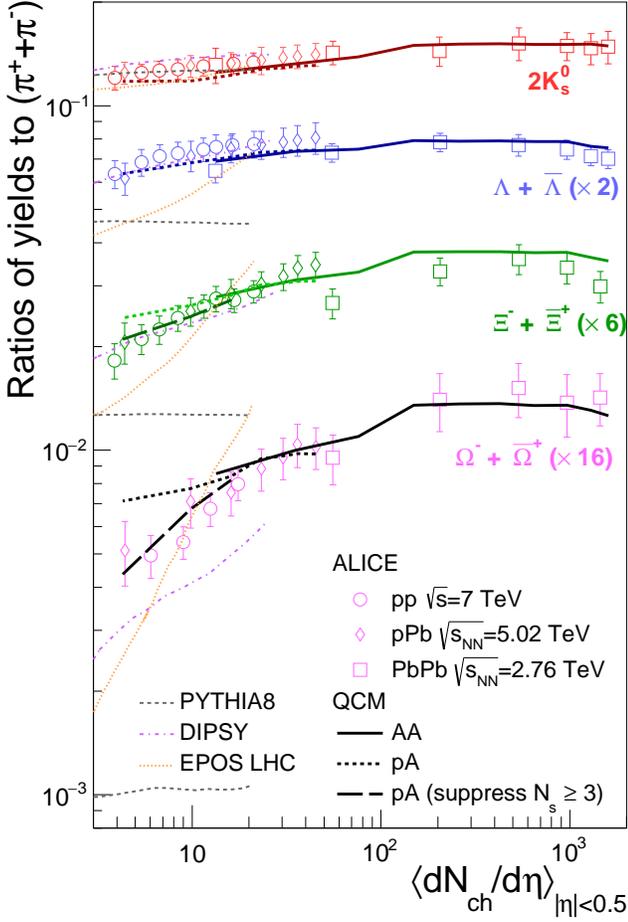}\\
	\caption{(Color online) Yield ratios of strange and multi-strange hadrons to pions as the function of charged-particle multiplicity in the pseudo-rapidity interval $|\eta|<0.5$. The experimental data of p+p, p+Pb, and Pb+Pb collisions are from Refs.~\cite{KLamXiOmgpp7TeV,piKpLampPb,xiomepPb,piKpPbPb,Ks0LPbPb,XiOmgPbPb}. Results of PYTHIA8, EPOS and DIPSY are taken from Ref.~\cite{KLamXiOmgpp7TeV}.}
  \label{fig2}
\end{figure}

In Fig.~\ref{fig2}, we show yield ratios of $K_s^0$, $\Lambda+\bar{\Lambda}$, $\Xi^{-}+\bar{\Xi}^{+}$ and $\Omega+\bar{\Omega}^{+}$ to $\pi^{+} + \pi^{-}$ as the function of charged particle multiplicity density at mid-rapidity. Symbols are the experimental data of p+p collisions at $\sqrt{s}=7$ TeV \cite{KLamXiOmgpp7TeV}, p+Pb collisions at $\sqrt{s_{NN}}=5.02$ TeV \cite{piKpLampPb,xiomepPb} and Pb+Pb collisions at $\sqrt{s_{NN}}=2.76$ TeV \cite{piKpPbPb,Ks0LPbPb,XiOmgPbPb}. Data of p+p and p+Pb collisions show an enhanced production of strange hadrons with the increasing charged particle multiplicity density, and at high multiplicity events they tend to be consistent with the data of Pb+Pb collisions within the statistical and systematic uncertainties. The enhancement of multi-strange baryons $\Xi^-$ and $\Omega^-$ are more pronounced than that of single strange hadrons like kaon and $\Lambda$. Such a species-dependent enhancement can be used to study the microscopic mechanism of hadron production at hadronization. 

In our model, yields of these strange and multi-strange hadrons relative to pions are mainly affected by $\lambda_s$ and $R_{B/M}$ but no longer by $\langle N_q \rangle$. Using Eqs.~(\ref{nbi_approx}) and (\ref{nmi_approx}), the hierarchy structure in the four yield ratios $K_s^0/\pi \equiv {2K_s^0}/(\pi^+ +\pi^-)$, $\Lambda/\pi \equiv (\Lambda + \bar{\Lambda} )/(\pi^+ +\pi^-)$, $\Xi/\pi \equiv (\Xi^- +\bar{\Xi}^+)/(\pi^+ +\pi^-)$, and $\Omega/\pi \equiv (\Omega^- + \bar{\Omega}^+ )/(\pi^+ + \pi^-)$ can be understood approximately 
\begin{eqnarray}
	\frac{K_s^0}{\pi} &\approx& \frac{  \lambda_s + 0.296\lambda_s^2 }{ 2.43+0.8\lambda_s+0.292\lambda_s^2 + \frac{\frac{8}{3}+2.024\lambda_s + \frac{4}{3}\lambda_s^2}{2+\lambda_s} R_{B/M}  },  \\
	\frac{\Lambda}{\pi} &\approx& \frac{ 7.736\lambda_s  }{  \frac{(2+\lambda_s) ( 2.43+0.8\lambda_s+0.292\lambda_s^2)}{R_{B/M}}  +  ( \frac{8}{3}+2.024\lambda_s + \frac{4}{3}\lambda_s^2)      }, \\
	\frac{\Xi}{\pi} &\approx& \frac{ 3\lambda_s^2  }{  \frac{(2+\lambda_s) ( 2.43+0.8\lambda_s+0.292\lambda_s^2)}{R_{B/M}}  +  ( \frac{8}{3}+2.024\lambda_s + \frac{4}{3}\lambda_s^2)      },  \\
    \frac{\Omega}{\pi} &\approx& \frac{ \lambda_s^3  }{  \frac{(2+\lambda_s) ( 2.43+0.8\lambda_s+0.292\lambda_s^2)}{R_{B/M}}  +  ( \frac{8}{3}+2.024\lambda_s + \frac{4}{3}\lambda_s^2)      }.
\end{eqnarray}
With parameter values $\lambda_s \sim 0.34 - 0.42$ and $R_{B/M} \sim 1/11 - 1/12$ as listed in Table \ref{tab1}, four yield ratios are 0.12-0.15, 0.035-0.038, 0.0044-0.0060, 0.00049-0.00080, respectively.
These analytic results are very close to the full results using formulas in Sec.~II, and well reproduce the observed hierarchy structure among four yield ratios shown in Fig.~\ref{fig2}.

The thick solid and dashed lines are our full results using formula in Sec.~II. As $\langle \mathrm{d}N_{ch}/\mathrm{d}\eta \rangle \gtrsim 20$ we can see that results of QCM on four yield ratios are all consistent with the experimental data.  In small $\langle \mathrm{d}N_{ch}/\mathrm{d}\eta \rangle \lesssim 20$ events, our results of $\Lambda$ and $K_s^0$ (the thick dashed lines) are still consistent with the data but those of $\Xi$ and $\Omega$ are higher than the data. 
We note that, for $\langle \mathrm{d}N_{ch}/\mathrm{d}\eta \rangle \lesssim 20$, the average strange quark number in mid pseudorapidity region $|\eta|<0.5$, denoted as $\langle N_s \rangle$, is smaller than three, in which case the threshold effect will appear for multistrange hadron production.
Taking $\Omega^-$ production as an example, only events with $N_s \geq 3$ have the positive probability of $\Omega^-$ formation while events $N_s <3$ are forbidden. In our calculations, we adopt Poisson distribution to approximate the distribution of strange quark number which has a long tail in large $N_s$ side. This might give too high population for events with $N_s \geq3$ and relate to overestimations of $\Xi$ and $\Omega$ yields.
In order to exclude the possible bias caused by the improper quark number distribution which might influence our final understanding of hadron production mechanism at LHC, we test other distribution shape for strange quarks based on Poisson distribution. 
We suppress the Poisson tail by a piece-wise function $\Theta(N_s)=\left\{ \left\{1,N_s<3\right\}, \left\{a,N_s\geq 3\right\}\right\} $ with a suppression parameter $a\leq1$. The new distribution is $P(N_s;\langle N_s \rangle) = \mathcal{N}\, Pois( N_s; \mu ) \Theta(N_s)$ where $\mathcal{N}$ is the normalization factor and $\mu$ is solved by the average constraint $\sum_{N_s} P(N_s;\langle N_s \rangle)\, N_s = \langle N_s \rangle $ at a specific value $a$. By fitting the data of peripheral p+Pb collisions, we get three values of $a$, 0.6, 0.8 and 0.88 in centralities $80-100\%$, $60-80\%$ and $40-60\%$, respectively.
The thick long-dashed lines are our results with suppressed $N_s \geq 3$ events, and we find that both the data of $\Xi$ and those of $\Omega$ can be well reproduced simultaneously. Results of kaon and $\Lambda$ with suppressed $N_s \geq 3$ events are almost the same as those without suppression and are not shown in Fig.~\ref{fig2} for clarity. Therefore, we argue that the production of kaon, $\Lambda$, $\Xi$, and $\Omega$ in small collision systems (p+p and p+Pb) and in large collision system (Pb+Pb) at LHC could be consistently understood with the quark combination mechanism together with a better understanding of the fluctuations (especially the strangeness fluctuation). 

Results of other models are also shown in Fig.~\ref{fig2}, which are taken from Ref.~\cite{KLamXiOmgpp7TeV}. PYTHIA8 (thin dashed lines) can explain the kaon yields in small system created in p+p and p+Pb collisions, but under-estimate the relative yield of $\Lambda$ and significantly under-estimate the relative yields of $\Xi$ and $\Omega$. In addition, it predicts almost constant production fraction for strange baryons, which is contrary to the experimental data.  DIPSY has considered the possible interaction between strings and thus color ropes are allowed to form to give the increased production of baryons and strangeness. DIPSY (thin dashed dotted lines) can explain the relative production of kaon, $\Lambda$ and $\Xi$ but still under-predict that of $\Omega$. EPOS (thin dotted lines) can explain the kaon production but predict a too rapid increase for the fraction of strange baryons with respect to the charged particle multiplicity density. 

\section{summary }
We have used the quark combination model to study the yield ratios of identified hadrons produced in high multiplicity p+p, p+Pb and Pb+Pb collisions at LHC.
We discussed two classes of yield ratios which can reflect two important features of hadronization. The first is the $p/\pi$ and $\Lambda/K_s^0$ yield ratios which can reflect the probability of baryon production at hadronization, and the second is $K_s^0/\pi$, $\Lambda/\pi$, $\Xi/\pi$ and $\Omega/\pi$ ratios relating to strangeness production. 
The experimental data of $p/\pi$ and $\Lambda/K_s^0$ can be reproduced simultaneously by the quark combination, and our results suggest the same probability of baryon production in light and strange sectors which can be inferred from the flavor independence of strong interactions. Moreover, the experimental data of $p/\pi$ and $\Lambda/K_s^0$ in high multiplicity p+p, p+Pb collisions are consistent with those in Pb+Pb collisions and two ratios are little changed over three orders of magnitude in charged particle multiplicity. This is a strong indication of the universal mechanism in baryon production for different quark gluon final states in three collision systems at LHC.
The data of $K_s^0/\pi$, $\Lambda/\pi$, $\Xi/\pi$ and $\Omega/\pi$ show a hierarchy property relating to the strangeness content, and they are naturally explained by quark combination in a simple analytic way.
We find that for small systems $\langle \mathrm{d}N_{ch}/\mathrm{d}\eta \rangle \lesssim 20$ in which strange quark number is $\langle N_s \rangle \lesssim 3$ the threshold effect of mutli-strange baryons (e.g.,~$\Omega^-$) is significant, and therefore the fluctuation of strangeness production is important for the production of strange hadron and, in particular, mutli-strange hadrons in small system.

\section*{Acknowledgements}

The authors thank Gang Li and Lie-wen Chen for helpful discussions.
This work is supported in part by the National Natural Science Foundation of China under Grant Nos. 11575100, 11675091, 11505104 and 11305076.

\end{document}